\documentclass[
reprint,
superscriptaddress,
amsmath,amssymb,
aps,
]{revtex4-2}

\usepackage{graphicx}% Include figure files
\usepackage{dcolumn}% Align table columns on decimal point
\usepackage{bm}% bold math
\usepackage[utf8]{inputenc}
\usepackage[T1]{fontenc}
\begin{document}
	
	\preprint{APS/123-QED}
	
	\title{Reconfigurable non-reciprocal wave growth in spatiotemporal modulated  1-D crystal }
	
	\author{Mohamed F. Hagag}
	
	\affiliation{Electronic Engineering Department$,$ Military Technical College$,$ Cairo  11766$,$ Egypt}
	\affiliation{Electronics and Communications Engineering Department$,$ American  University  in Cairo$,$ Cairo   11835$,$  Egypt}
	
	\author{Thomas R. Jones}%
	\affiliation{Elmore Family School of Electrical and Computer Engineering$,$  Purdue   University$,$ West  Lafayette$,$ Indiana  47907$,$  USA}
	
	\author{Karim Seddik}%
	\affiliation{Electronics and Communications Engineering Department$,$  American  University   in  Cairo$,$  Cairo   11835$,$  Egypt}
	\author{Dimitrios Peroulis}
	
	\affiliation{Elmore Family School of Electrical and Computer Engineering$,$ Purdue University$,$ West  Lafayette$,$  Indiana   47907$,$ USA}
 
	
	\begin{abstract}
		%We investigate nonreciprocal signal amplification in a space-time-modulated transmission line. By loading a transmission line with a sinusoidally time-modulated capacitor, reciprocal momentum band gaps (MPG) appear in the unit cell dispersion diagram (DD) at frequency ratio $F_{signal}= 0.5~F_{modulation}$. Due to the sole existence of complex frequencies in the MPG, waves propagating with a momentum that lies within the MPG are amplified. In addition to time modulation, the TL is space-modulated by imposing a modulation phase shift $-\theta_m$ between successive unit cells. As a result, forward and backward MPG locations in the DD shift from $F_{signal}= 0.5~F_{modulation}$ in opposite directions, causing nonreciprocal signal growth frequencies. Nonreciprocal amplification is confirmed by circuit modeling outcomes, which totally coincide with DD results.    
		Nonreciprocity in space-time modulated photonic crystals has been investigated in the context of nonreciprocal propagation and polarization. Here, we investigate a reconfigurable nonreciprocal wave growth in space-time modulated crystals. Imposing an adaptable progressive phase shift between successive time-modulated cells results in blue and red shifts of the forward and backward momentum band gaps around the typical 0.5 growth normalized frequency. We applied this spatiotemporal scheme to engineering the dispersion relation of a loaded transmission line$-$a 1D periodic structure$-$in the microwave regime.
		%\begin{description}
		%\end{description}
	\end{abstract}
	
	\maketitle
	
	%\tableofcontents
	
	%\section{Introduction}
	\textit{Introduction.}$-$ Growing wave physics gains high interest due to the continuous demand for high-power signals in all frequency regimes. Creating sufficient nonlinearities and instabilities in a media allows waves to grow due to wave-matter energy coupling \cite{PhysRev.112.1488}. It is first started with the interaction between plasma and charged-particle beams, which is utilized to generate high-power microwave signals \cite{osti_4815657,1964}. Another approach is the parametric amplification that allows wave growth by pumping a system with a high-amplitude pump, coupling the energy to a traveling signal \cite{CULLEN1958,Tien1958, Huang2018}. Recently, utilizing time as an extra degree of freedom has created new types of artificial electromagnetic media, generally called time crystals (TC). Periodically modulating media parameters allows the existence of unusual physical effects such as momentum gaps, magnetless non-reciprocity \cite{Yu2009,Ruesink2016}, parity-time symmetry \cite{PhysRevLett.127.153903}, and topological aspects \cite{Lustig18,Wang2021}. Analog to photonic crystals (PC) in which the media is space-modulated, creating frequency band gaps \cite{Joannopoulos2008-mv,Sievenpiper1999HighimpedanceES}, momentum (k) band gaps (MPG) are created in the dispersion diagram of TCs if the media parameters are time-modulated with a sufficiently high speed. Only complex frequency exists within an MPG, allowing the waves to grow \cite{Lustig18,10.1117/1.AP.4.1.014002}. However, utilizing only time modulation, TC has reciprocal MPGs and amplifies forward and backward propagating waves only if the wave frequency is equal to half of the modulation frequency \cite{Galiffi2023, doi:10.1126/sciadv.adg7541}. 
 
Non-reciprocity is an exotic property of TC that can be achieved due to breaking time-reversal symmetry \cite{Sounas2017}. Spatiotemporal modulation of traveling waves purvey an artificial linear momentum to a structure. This imposed artificial momentum plays the main role of breaking that symmetry, causing non-reciprocity \cite{Sounas2017}. Over the last few years, this phenomenon has been utilized to realize many non-reciprocal propagation in magnetless devices, such as isolators,  phase shifters, transmission lines, and metamaterials \cite{Yu2009, Ruesink2016, 1425478, Hadad2015, 6887369}. In addition to non-reciprocity in wave propagation, the non-reciprocal gain is investigated by combining special effects with TC, or MPG, such as temporal Faraday effects \cite{He2023}. The ability to synthesize an artificial linear momentum combined with MPGs in spatiotemporal TC gives rise to a new approach to engineering a reconfigurable non-reciprocal gain in periodic crystals. In this work, we research a reconfigurable non-reciprocal wave growth in a spatiotemporal modulated 1-D microwave periodic crystal. A loaded TL consists of T-shape unit cells with a time-invariant TL as a series element and a time-modulated capacitor (TMC) with a sinusoidal waveform as a shunt element. The TL is space-modulated by forcing an adaptable progressive phase shift $-\theta_m$ between successive cells. The eigenvalue problem is solved, and the dispersion diagram is plotted at different values of $\theta_m$, showing MPGs' different locations for forward and backward propagation. To confirm the nonreciprocal behavior, the space-time modulated TL is transient simulated using circuit modeling. The simulation results show nonreciprocal amplification frequencies that align with the eigenvalue problem solution.

\begin{figure*}[!t]
		\centering
		\includegraphics[width=.95\linewidth]{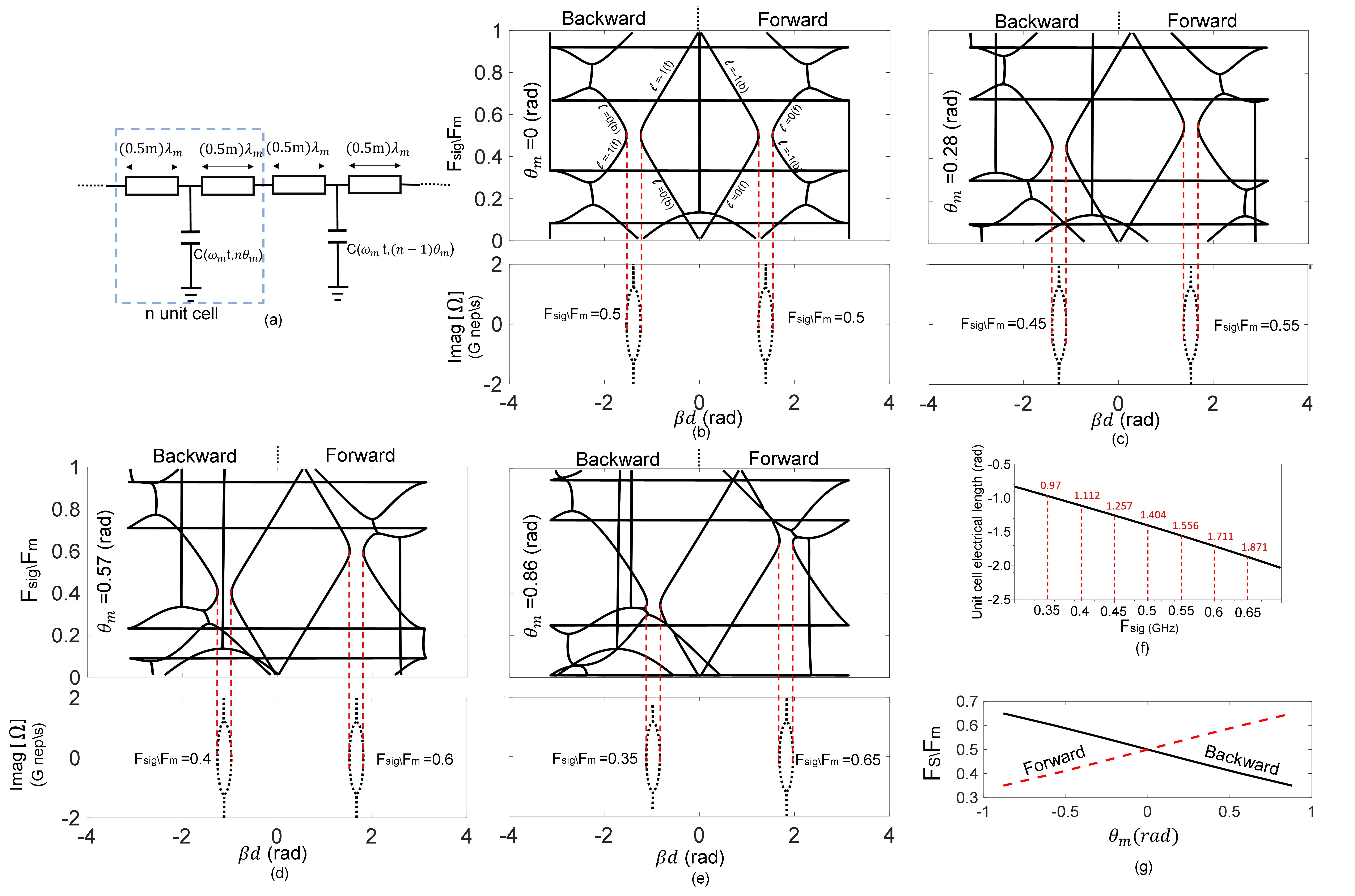}
		\caption{ (a) T-shape unit cell with a time-invariant TL of length $0.5m~\lambda_m$ as series element and space-time modulated capacitor as shunt element. Real and complex frequency dispersion diagram with $N=2$ for the unit cell in (a) with $m=0.3$, $C_o=~4~$pF, $M_D=0.7$, $F_m= 1$~GHz and TL $Z_o=$ $83~\Omega$ at (b) $\theta_m=~0$ rad, (c) $\theta_m=~0.28$ rad, (d) $\theta_m=~0.57$ rad and (e) $\theta_m=~0.86$ rad. (f) Unit cell electrical length variation with frequency at $C_o$. (g) Signal growth frequency variation for forward and backward propagation with $\theta_m$. (In Fig.\ref{DD}(b) $l$ is the order of harmonic with (f) and (b) referring to forward and backward propagation, respectively.) }
		\label{DD}
	\end{figure*}

	\textit{Dispersion relation.}$-$ The considered loaded transmission line (TL) unit cell is shown in Fig. \ref{DD}(a). It is a T-shape unit cell with two time-invariant (TI) TLs of length $0.5m~\lambda_m$, where $\lambda_m$ is the wavelength of the modulation signal, as series elements and a TMC as a shunt element. The TMC is modulated in time periodically with the period $T_m$ following the function $C(\omega,t)=C(\omega,t+lT_m)\ $ with $ l\in\mathbb{Z}$. Moreover, it is also space-modulated by adding a progressive modulation phase shift $-\theta_m$ in successive cells. Following \cite{9063633,Jayathurathnage2021,Elnaggar2021} and considering a constant phase shift $-\theta_m$ in the modulation of TMC between sequential unit cells, the space-TMC can be expanded into a complex Fourier series as follows.
	\begin{equation}
	%C\left(\omega t,\right)=\sum_{s=-\infty}^{+\infty}{c_s(\omega)e^{js\omega_Mt}}
	%C\left(\omega t, \theta_m\right)=\sum_{l=-\infty}^{+\infty}\sum_{s=-\infty}^{+\infty}{c_s(\omega)e^{js\omega_mt}e^{-jl\theta_m}}
	C\left(\omega_m t, \theta_m\right)=\sum_{s=-\infty}^{+\infty}{c_s(\omega)e^{js\omega_mt}e^{-js\theta_m}}
	\label{eq:C}
	\end{equation}
	where $c_s(\omega)$ are complex coefficients and $\omega_m$ is the angular modulation frequency. A system with a TMC will possess an infinite number of Floquet harmonics that follows 
	\begin{equation}
	\omega_l=\omega_{sig}+l\omega_m
	\label{eq:om}
	\end{equation}
	where $\omega_{sig}$ is the main signal angular frequency and  $l$ is the Floquet order. Across the TMC,  the voltage or current follows the series below \cite{Jayathurathnage2021}
	
	\begin{equation}
	F\left(t\right)=\sum_{l=-N}^{+N}{f_le^{j\omega_lt}}
	\label{eq:V}
	\end{equation}
	where $f_l$ are the complex coefficients of voltage and current. $N$ is an integer that represents half the number of considered harmonics. As a result, the total current can be found in terms of voltage harmonics as follows  
	\begin{multline}
	I\left(\omega t,\theta_m\right)=\sum_{l=-\infty}^{\infty}\sum_{s=-\infty}^{\infty}{j~\left(\omega_l+s\omega_m\right)~c_s(}\omega_l)\\
	\times e^{js\omega_mt}e^{-js\theta_m}~v_l~e^{j\omega_lt}
	\label{eq:N}
	\end{multline}
	The relation between current and voltage harmonics can be found using (\ref{eq:V}) and (\ref{eq:N}) as follows.  
	
	\begin{multline}
	\sum_{r=-\infty}^{\infty}{i_re^{jr\omega_lt}}=\sum_{l=-\infty}^{\infty}\sum_{s=-\infty}^{\infty}{j~\omega_{l+s}~c_s(}\omega_l)\\
	\times e^{-js\theta_m}~v_l~e^{j\omega_{l+s}t}
	\label{eq:N1}
	\end{multline}
	By equating the index $ r $ by $l+s$ and matching $\omega_r$ frequency, we get
	\begin{equation}
	i_r=\sum_{l=-\infty}^{\infty}{j~\omega_r~c_{r-l}(\omega_{l})~v_{l}~e^{-j(r-l)\theta_m}}
	\label{eq:N2}
	\end{equation}
	Consequently, due to space modulation, the relation between the Y-matrix elements in two successive  cells $n$ and $n+1$ is given by
	\begin{equation}
	{Y_{(n+1)}}_r^l={Y_{(n)}}_r^l~e^{-i(r-l)\theta_m} 
	\label{eq:YC_e}
	\end{equation} 
	Hence, the total impedance matrix is given by   
	\begin{equation}
	\bar{\bar{Y_C}}=j\bar{\bar{W}}P~\bar{\bar{\xi}}
	\label{eq:YC}
	\end{equation}
	where $\bar{\bar{W}}=diag(\omega_{-N},..,\omega_{-1},\omega_s,\omega_{1},..,\omega_N)$, $\bar{\bar{\xi}}=diag(e^{-j(-N)\theta_m},..,e^{-j(-1)\theta_m},1,e^{-j(1)\theta_m},..,e^{-j(N)\theta_m})$. $P$ is  $(2N+1) \times (2N+1)$ in size which is mainly dependent on the modulation waveform. Considering TMC with sinusoidal modulation that follows
	\begin{equation}
	C\left(\omega,t\right)=C_o(\omega)\left( 1+M_D\cos{\omega_mt}\right)
	\label{eq:C1}
	\end{equation}
	where $M_D$ is the modulation depth and  $C_o(\omega)$ is the capacitance nominal value. $P$ is given by
	\begin{multline}
	P=\\
	\left(\begin{matrix}\begin{matrix}c_0(\omega_{-N})&c_{-1}(\omega_{1-N})&0\\c_1(\omega_{-N})&c_0(\omega_{1-N})&c_{-1}(\omega_{2-N})\\0&c_1(\omega_{1-N})&c_0(\omega_{2-N})\\\end{matrix}&\begin{matrix}\cdots\\\cdots\\\cdots\\\end{matrix}&\begin{matrix}0\\0\\0\\\end{matrix}\\\begin{matrix}\vdots&\ \ \ \ \ \ \ \ \ \ \ \ \ \ \ \ \vdots&\ \ \ \ \ \ \ \ \ \ \ \ \ \ \ \ \ddots\\\end{matrix}&\ddots&\vdots\\\begin{matrix}0\ \ \ \ \ \ \ \ \ \ \ \ \ \ \ &0\ \ \ \ \ \ \ \ \ \ \ \ \ \ \ \ &0\\\end{matrix}&\cdots&c_0(\omega_N)\\\end{matrix}\right)
	\label{eq:YC1}
	\end{multline}
	where $c_{\pm1}(\omega){=0.5~c_o(\omega) \ M_D}$. It is worth mentioning that considering only time modulation, $\theta_m=0$, the admittance matrix of TMC will be reduced to
 \begin{equation}
	\bar{\bar{Y_C}}=j\bar{\bar{W}}P
	\label{eq:YCt}
	\end{equation}
	According to \cite{Elnaggar2021}, the transfer matrix of the space-TMC, which has the admittance matrix shown in (\ref{eq:YC}), can be obtained by multiplying each element of the transfer matrix of the TMC, which has the admittance matrix shown in (\ref{eq:YCt}), by the space factor $\bar{\bar{\xi}}$. Consequently, the transfer matrix of a space-TMC can be obtained using
	\begin{equation}
	T_C=\left[\begin{matrix}\bar{\bar{Ones}}&\bar{\bar{Zeros}}\\j\bar{\bar{W}}P&\bar{\bar{Ones}}\\\end{matrix} \right] \times \left[\begin{matrix}\bar{\bar{\xi}}&\bar{\bar{Zeros}}\\\bar{\bar{Zeros}}&\bar{\bar{\xi}}\\\end{matrix}\right]
	\label{eq:TC}
	\end{equation}
	where $\bar{\bar{Ones}}$ and $\bar{\bar{Zeros}}$ are the unit and zero matrices, receptively. The transfer matrix of the TI TL of length $0.5~m~\lambda_m$ is given by
	
	\begin{equation}
	T_{tl}=\left[\begin{matrix}\cos(\bar{\bar{\varphi}})&j \times Z_o \times \sin (\bar{\bar{\varphi}})\\ j \times Y_o \times \sin (\bar{\bar{\varphi}} )&\cos(\bar{\bar{\varphi}})\\\end{matrix}\right]
	\label{eq:TTL}
	\end{equation}
	where  $Z_o$ and $Y_o$ are the TL characteristic impedance and admittance, respectively and $\bar{\bar{\varphi}}$ is given by
	\begin{equation}
	\bar{\bar{\varphi}}= 2\times \pi \times m \times \frac{\bar{\bar{W}}}{\omega_m}
	\label{eq:TTL1}
	\end{equation}
	
	The total transfer matrix can be obtained as follows.
	\begin{equation}
	T_t= T_{tl} \times T_C \times T_{tl}
	\label{eq:TT}
	\end{equation}
	Finally, the dispersion diagram (DD) can be plotted by solving 
	\begin{equation}
	T_t X=e^{ \gamma d} X
	\label{eq:dd}
	\end{equation}
	where $\gamma$, $X$, and $d$ are the propagation constant, the eigenvectors, and the unit cell (UC) length, respectively. $\gamma$ is related to phase constant $\beta$ and  the attenuation constant $\alpha$ by
	\begin{equation}
	\gamma=\alpha +j \beta
	\label{eq:dd1}
	\end{equation}
	
	%\section{Signal growth in Non-reciprocal Dispersion diagram  of Space-time modulated TL}\label{S2}
	
	\textit{Wave growth in nonreciprocal dispersion diagram of space-time modulated TL.}$-$ For the unit cell shown in Fig. \ref{DD}(a), the Bloch impedance ($Z_{Bloch}$) of the unit cell is chosen to be 50 $\Omega$ at $F_{sig}= 0.5$ GHz. As result, $C_o$, TL characteristic impedance $Z_o$ and $m$ are set to $4$~ pF, $83~\Omega$ and $0.3$, respectively.  The dispersion diagram (DD) is plotted in Fig. \ref{DD}(b) considering only time modulation ($F_m= 1$~GHz and $\theta_m=~0$) and $M_D=0.7$ with $N=2$. At $F_{sig}= 0.5$~GHz ($F_m= 2F_{sig}$), a phase matching occurs between  the  fundamental $\omega_{sig}$ and the harmonic $\omega_{-1}$ in forward and backward propagations. The matching phase equals the unit cell electrical length at $C_o$ ($UCEL_{C_o}$). Consequently, strong interaction happens, and momentum band gaps (MBG) for forward and backward propagation are formed, in which no relation between real frequency and phase exists \cite{Lustig18,10.1117/1.AP.4.1.014002}. As shown in Fig. \ref{DD}(b), within the MBG, only complex frequency ($\omega_{sig}$ and $\omega_{-1}$) exists with a constant real part equal $F_{sig}= 0.5$~GHz ($F_m= 2F_{sig}$) and varying imaginary part. The imaginary part has positive values (causing wave growth) and negative values (causing wave decay) with a maximum centered at the middle of the gap. This behavior is similar to the exceptional point physics \cite{RevModPhys.93.015005,Miri2019}. The complex frequency dispersion relation is plotted using the following equation instead of (\ref{eq:om}) 
	\begin{equation}
	\Omega=\omega_l-i\sigma_l
	\label{eq:omc}
	\end{equation} 
	
	Next, in addition to time modulation, space modulation is considered by imposing a progressive phase shift ($-\theta_m$) to the modulation between successive cells.  
 we will focus on the substantial harmonics $\omega_{0}$ and $\omega_{-1}$ that are linked to the MPGs. According to (\ref{eq:TC}), the dispersion lines of fundamental $\omega_{sig}$ are not affected. On the other hand, the forward and backward dispersion lines of harmonic $\omega_{-1}$ are shifted in a way that causes the forward and backward MPGs to be settled at two different frequencies following
	\begin{equation}
	{UCEL_{C_o}\ at\ \left(F_{f}\right)}-{UCEL_{C_o}\ at\ \left(F_{b}\right)}= \theta_m
	\label{eq:f}
	\end{equation}  
	where $F_{f}$ and $F_{b}$ are the frequencies at which MBGs are formed at forward and backward propagation, respectively. In addition, $F_{f}$ and $F_{b}$ are related as
	\begin{equation}
	\frac{F_{f}+F_{b}}{F_m}=1
	\label{eq:f1}
	\end{equation} 
	
	\begin{figure*}[!t]
		\centering
		\includegraphics[width=.8\linewidth]{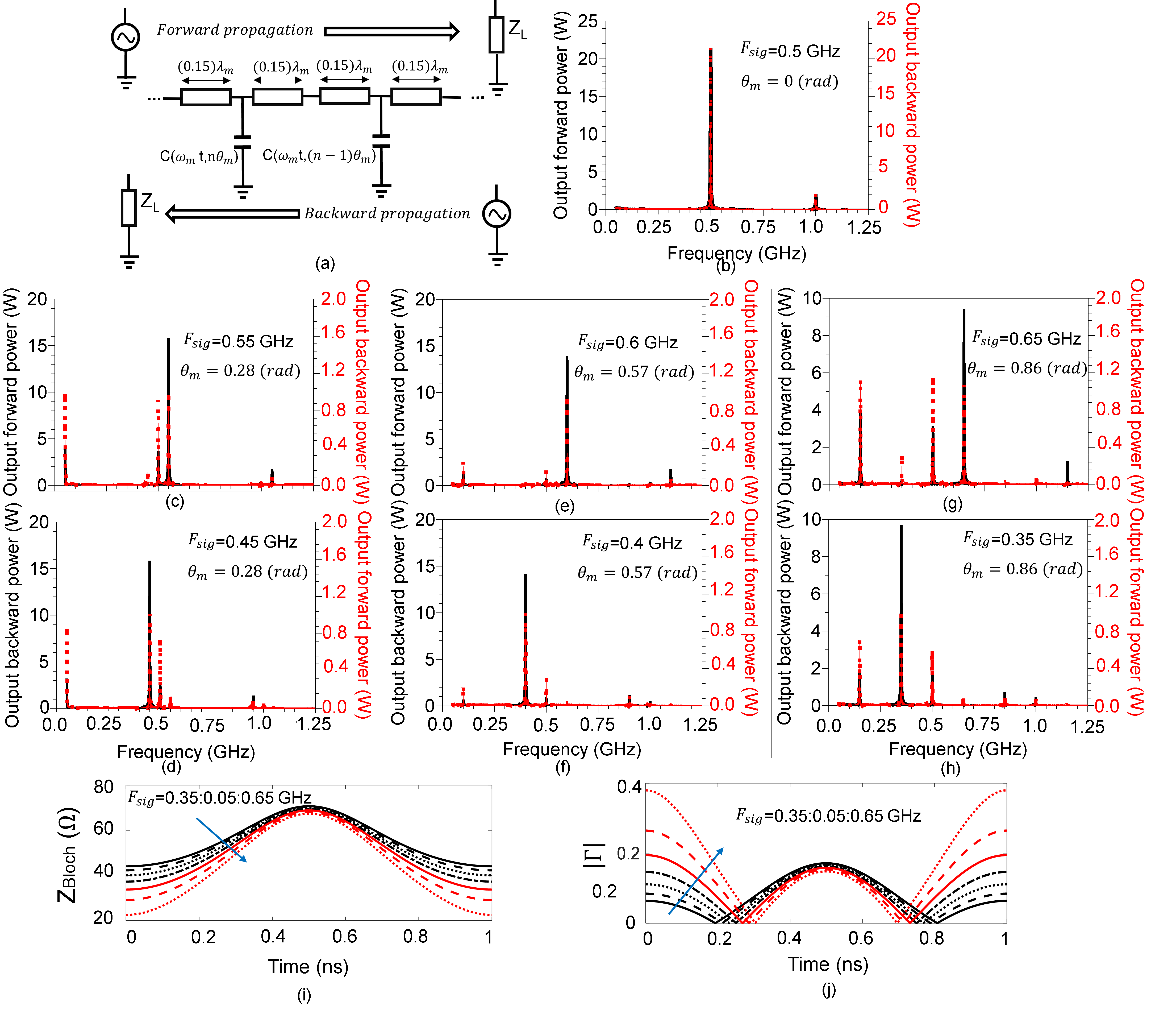}
		\caption{(a) Simulated nine unit cells with 1 W input power, TLs of length $0.15\lambda_m$, $Z_o=83~\Omega$, $C_o =4$~pF, $F_m =1$~GHz, $M_D=0.7$ loaded with a $50~\Omega$ load impedance. Forward (backward) propagation is defined in the direction where the phase shift between successive cells is $-\theta_m$ ($\theta_m$). Output forward and backward power in case (b) input frequency $F_{sig}=0.5$~GHz and $\theta_m=0$ rad, (c) $F_{sig}=0.55$~GHz and $\theta_m=0.28$ rad, (d) $F_{sig}=0.45$~GHz and $\theta_m=0.28$ rad, (e) $F_{sig}=0.6$~GHz and $\theta_m=0.57$ rad,  (f) $F_{sig}=0.4$~GHz and $\theta_m=0.57$ rad,  (g) $F_{sig}=0.65$~GHz and $\theta_m=0.86$ rad, (h) $F_{sig}=0.35$~GHz and $\theta_m=0.86$ rad. At different values of $F_{sig}$ and with time dependence, (i) unit cell $Z_{Bloch}$, (j) reflection coefficient considering $50~\Omega$ load and source.}
		\label{ADS}
	\end{figure*}
	
	Equation (\ref{eq:f}) describes the required value of $\theta_m$ that makes harmonics $\omega_{0}$ and $\omega_{-1}$ match phases at $C_o$ at specific frequencies. It works precisely to define wave growth frequencies at low values of $M_D$; however, as $M_D$ increases, required $\theta_m$ deviates slightly from (\ref{eq:f}) and can be easily optimized. This happens because the unit cell has time-invariant elements, TLs. As a result, the unit cell $Z_{Bloch}$ does not follow the capacitance variation in a linear behavior. Hence, considering only time modulation, harmonics $\omega_{0}$ and $\omega_{-1}$ matching phases at $C_o$  shift from the MPG center. When spatiotemporal modulating the line, wave growth frequencies shift slightly from the frequencies at which harmonics $\omega_{0}$ and $\omega_{-1}$ match phases at $C_o$, and a minor optimization is needed. For further investigation, DDs are plotted for $\frac{F_f}{F_m}$ values $0.55$, $0.6$ and $0.65$ in Figs. \ref{DD}(c-e), respectively. As shown, $\frac{F_{b}}{F_m}$ exactly follows (\ref{eq:f1}). On the other hand, the calculated electrical lengths of the unit cell at different frequencies, shown in Fig. \ref{DD}(f), are utilized in (\ref{eq:f}) to obtain the initial value of $\theta_m$ and then optimized. At the exact value of $\theta_m$ associated with a desired $\frac{F_f}{F_m}$ ($\frac{F_{b}}{F_m}$), complex frequencies exist with a varying imaginary part and fixed real part defined by the desired $\frac{F_f}{F_m}$ ($\frac{F_{b}}{F_m}$). For $\frac{F_f}{F_m}$ ($\frac{F_b}{F_m}$) values $0.55$ ($0.45$), $0.6$ ($0.4$) and $0.65$ ($0.35$), $\theta_m$ optimized values are found to be $0.28$, $0.57$ and $0.86$ rad, respectively. Moreover, as shown in Figs. \ref{DD}(b-e), imaginary part values of $\Omega$ are almost the same at different values of $\theta_m$. Consequently, the same wave growth rate is expected if the same boundary conditions exist. According to Fig. \ref{DD}(g), $\frac{F_f}{F_m}$ and  $\frac{F_{b}}{F_m}$ can take variety of values lower or higher than $0.5$ by choosing the proper value and sign of $\theta_m$, (\ref{eq:f1}) is always applied. 
	
	%\section{Circuit Modeling}
	
	\textit{Circuit Modeling.}$-$ Scenarios presented in Figs.\ref{DD}(b-e) are verified using transient simulations (TS) and circuit modeling. As illustrated in Fig. \ref{ADS}(a), nine-unit cells are considered with 1 W input power source followed by $50~\Omega$ isolator and $50~\Omega$ load impedance. TLs have a length of $0.15\lambda_m$ and $Z_o=83~\Omega$. The capacitor has $C_o =4$~pF and time-modulated with $F_m =1$~GHz and $M_D=0.7$. Output forward (backward) power is the output power when the source and the load are connected to the terminals with a $-\theta_m$ ($\theta_m$) phase shift between successive cells, as shown in Fig. \ref{ADS}(a). In all considered cases, TS output results are Fourier transformed and plotted in Figs. \ref{ADS}(b-h).
	
	For input frequency ($F_{sig}$) equals $0.5$~GHz and $\theta_m=0$ rad, as shown in Fig. \ref{ADS}(b), the output forward power (OFP) and output backward power (OBP) are identical with a gain more than $13$ dB, congruent with Figs. \ref{DD}(b). Moreover, following Figs. \ref{DD}(c) and considering $F_{sig}=0.55$ ($0.45$) GHz with $\theta_m=0.28$ rad, OFP (OBP) is amplified to $11.7$ dB while no gain is achieved in OBP (OFP) as illustrated in Figs. \ref{ADS}(c-d). Finally, by following Figs.~\ref{DD}(d)/(e) and considering $F_{sig}=0.6$/$0.65$ ($0.4$/$0.35$)~GHz with $\theta_m=0.57$/$0.86$ rad, OFP (OBP) is amplified to $11.4$/$10$ dB without any gain in OBP (OFP) as shown in Figs. \ref{ADS}(e-f)/(g-h).
	
	As illustrated in Figs. \ref{DD}(b-e), when space-time modulating the TL, it is expected to have the same growth rate if the boundary conditions are the same. In the simulations illustrated in Fig. \ref{ADS}, loaded TL is terminated in all cases with a fixed $50~\Omega$ impedance at the source and the load. The TL is matched at $C_o$ and $F_{sig}=0.5$~GHz to a fixed $50~\Omega$ impedance. However, shifting from  $F_{sig}=0.5$~GHz changes the $Z_{Bloch}$ profile, Fig. \ref{ADS}(i), with modulation period (T$_m$) which changes the reflection coefficient profile at the terminals, Figs. \ref{ADS}(j). Consequently, the boundary conditions change, resulting in a growth rate variation shown in Fig. \ref{ADS}. However, studying the detailed effect of boundary conditions variation on amplification gain is beyond the scope of this letter.    
	
	%\section{Conclusion}
	
	In conclusion, reconfigurable nonreciprocal wave growth in a space-time modulated TL is confirmed through eigenvalue problem solution and circuit modeling. TL is time-modulated by loading sinusoidally time-modulated capacitors and space-modulated by forcing a modulation phase shift $-\theta_m$ between successive cells. Varying $\theta_m$ causes MPGs to shift from $F_{signal}= 0.5~F_{modulation}$ in a nonreciprocal behavior. As a result, nonreciprocal wave growth frequencies are obtained.   
 
	%\begin{acknowledgments}
This work has been completed under a research agreement between Purdue University and The American University in Cairo.
%\end
%\section*{Author Contributions}
%\textbf{Mohamed Hagag:} Conceptualization (lead); Data curation (lead); Formal analysis (lead); Investigation (equal); Methodology (lead); Validation (lead); Visualization (lead); Writing - original draft (lead). \textbf{Thomas Jones:} Data curation (equal); Investigation (equal); Resources (equal); Writing - review $\&$ editing (equal). \textbf{Karim Seddik:} Project administration (equal); Resources (equal); Supervision (equal); Writing - review $\&$ editing (equal). \textbf{Dimitrios Peroulis:} Funding acquisition (lead);; Investigation (lead); Project administration (lead); Resources (lead); Supervision (lead); Writing - review $\&$ editing (lead).
%\section*{Data Availability}
%The datasets used and/or analyzed during the current study are available from the corresponding author upon reasonable request.% Pr% Produces the bibliography via BibTeX.
%\section*{References}
\bibliography{bibfile1}% Pr% Produces the bibliography via BibTeX

\begin{thebibliography}{26}%
\makeatletter
\providecommand \@ifxundefined [1]{%
 \@ifx{#1\undefined}
}%
\providecommand \@ifnum [1]{%
 \ifnum #1\expandafter \@firstoftwo
 \else \expandafter \@secondoftwo
 \fi
}%
\providecommand \@ifx [1]{%
 \ifx #1\expandafter \@firstoftwo
 \else \expandafter \@secondoftwo
 \fi
}%
\providecommand \natexlab [1]{#1}%
\providecommand \enquote  [1]{``#1''}%
\providecommand \bibnamefont  [1]{#1}%
\providecommand \bibfnamefont [1]{#1}%
\providecommand \citenamefont [1]{#1}%
\providecommand \href@noop [0]{\@secondoftwo}%
\providecommand \href [0]{\begingroup \@sanitize@url \@href}%
\providecommand \@href[1]{\@@startlink{#1}\@@href}%
\providecommand \@@href[1]{\endgroup#1\@@endlink}%
\providecommand \@sanitize@url [0]{\catcode `\\12\catcode `\$12\catcode
  `\&12\catcode `\#12\catcode `\^12\catcode `\_12\catcode `\%12\relax}%
\providecommand \@@startlink[1]{}%
\providecommand \@@endlink[0]{}%
\providecommand \url  [0]{\begingroup\@sanitize@url \@url }%
\providecommand \@url [1]{\endgroup\@href {#1}{\urlprefix }}%
\providecommand \urlprefix  [0]{URL }%
\providecommand \Eprint [0]{\href }%
\providecommand \doibase [0]{https://doi.org/}%
\providecommand \selectlanguage [0]{\@gobble}%
\providecommand \bibinfo  [0]{\@secondoftwo}%
\providecommand \bibfield  [0]{\@secondoftwo}%
\providecommand \translation [1]{[#1]}%
\providecommand \BibitemOpen [0]{}%
\providecommand \bibitemStop [0]{}%
\providecommand \bibitemNoStop [0]{.\EOS\space}%
\providecommand \EOS [0]{\spacefactor3000\relax}%
\providecommand \BibitemShut  [1]{\csname bibitem#1\endcsname}%
\let\auto@bib@innerbib\@empty
%</preamble>
\bibitem [{\citenamefont {Sturrock}(1958)}]{PhysRev.112.1488}%
  \BibitemOpen
  \bibfield  {author} {\bibinfo {author} {\bibfnamefont {P.~A.}\ \bibnamefont
  {Sturrock}},\ }\bibfield  {title} {\bibinfo {title} {Kinematics of growing
  waves},\ }\href {https://doi.org/10.1103/PhysRev.112.1488} {\bibfield
  {journal} {\bibinfo  {journal} {Phys. Rev.}\ }\textbf {\bibinfo {volume}
  {112}},\ \bibinfo {pages} {1488} (\bibinfo {year} {1958})}\BibitemShut
  {NoStop}%
\bibitem [{\citenamefont {Fainberg}(1962)}]{osti_4815657}%
  \BibitemOpen
  \bibfield  {author} {\bibinfo {author} {\bibfnamefont {Y.~B.}\ \bibnamefont
  {Fainberg}},\ }\bibfield  {title} {\bibinfo {title} {Interaction of
  charged-particle beams with plasma},\ }\href
  {https://doi.org/10.1007/bf01473684} {\bibfield  {journal} {\bibinfo
  {journal} {The Soviet Journal of Atomic Energy}\ }\textbf {\bibinfo {volume}
  {11}},\ \bibinfo {pages} {958} (\bibinfo {year} {1962})}\BibitemShut
  {NoStop}%
\bibitem [{\citenamefont {Briggs}(1964)}]{1964}%
  \BibitemOpen
  \bibfield  {author} {\bibinfo {author} {\bibfnamefont {R.~J.}\ \bibnamefont
  {Briggs}},\ }\href {https://doi.org/10.7551/mitpress/2675.003.0001} {\emph
  {\bibinfo {title} {Electron-Stream Interaction with Plasmas}}}\ (\bibinfo
  {publisher} {The {MIT} Press},\ \bibinfo {year} {1964})\BibitemShut {NoStop}%
\bibitem [{\citenamefont {CULLEN}(1958)}]{CULLEN1958}%
  \BibitemOpen
  \bibfield  {author} {\bibinfo {author} {\bibfnamefont {A.~L.}\ \bibnamefont
  {CULLEN}},\ }\bibfield  {title} {\bibinfo {title} {A travelling-wave
  parametric amplifier},\ }\href {https://doi.org/10.1038/181332a0} {\bibfield
  {journal} {\bibinfo  {journal} {Nature}\ }\textbf {\bibinfo {volume} {181}},\
  \bibinfo {pages} {332} (\bibinfo {year} {1958})}\BibitemShut {NoStop}%
\bibitem [{\citenamefont {Tien}(1958)}]{Tien1958}%
  \BibitemOpen
  \bibfield  {author} {\bibinfo {author} {\bibfnamefont {P.~K.}\ \bibnamefont
  {Tien}},\ }\bibfield  {title} {\bibinfo {title} {Parametric amplification and
  frequency mixing in propagating circuits},\ }\href
  {https://doi.org/10.1063/1.1723440} {\bibfield  {journal} {\bibinfo
  {journal} {Journal of Applied Physics}\ }\textbf {\bibinfo {volume} {29}},\
  \bibinfo {pages} {1347} (\bibinfo {year} {1958})}\BibitemShut {NoStop}%
\bibitem [{\citenamefont {Huang}(2018)}]{Huang2018}%
  \BibitemOpen
  \bibfield  {author} {\bibinfo {author} {\bibfnamefont {J.}~\bibnamefont
  {Huang}},\ }\bibfield  {title} {\bibinfo {title} {Parametric amplifiers in
  optical communication systems: From fundamentals to applications},\ }in\
  \href {https://doi.org/10.5772/intechopen.73696} {\emph {\bibinfo {booktitle}
  {Optical Amplifiers - A Few Different Dimensions}}}\ (\bibinfo  {publisher}
  {{InTech}},\ \bibinfo {year} {2018})\BibitemShut {NoStop}%
\bibitem [{\citenamefont {Yu}\ and\ \citenamefont {Fan}(2009)}]{Yu2009}%
  \BibitemOpen
  \bibfield  {author} {\bibinfo {author} {\bibfnamefont {Z.}~\bibnamefont
  {Yu}}\ and\ \bibinfo {author} {\bibfnamefont {S.}~\bibnamefont {Fan}},\
  }\bibfield  {title} {\bibinfo {title} {Erratum: Complete optical isolation
  created by indirect interband photonic transitions},\ }\href
  {https://doi.org/10.1038/nphoton.2009.73} {\bibfield  {journal} {\bibinfo
  {journal} {Nature Photonics}\ }\textbf {\bibinfo {volume} {3}},\ \bibinfo
  {pages} {303} (\bibinfo {year} {2009})}\BibitemShut {NoStop}%
\bibitem [{\citenamefont {Ruesink}\ \emph {et~al.}(2016)\citenamefont
  {Ruesink}, \citenamefont {Miri}, \citenamefont {Al{\`{u}}},\ and\
  \citenamefont {Verhagen}}]{Ruesink2016}%
  \BibitemOpen
  \bibfield  {author} {\bibinfo {author} {\bibfnamefont {F.}~\bibnamefont
  {Ruesink}}, \bibinfo {author} {\bibfnamefont {M.-A.}\ \bibnamefont {Miri}},
  \bibinfo {author} {\bibfnamefont {A.}~\bibnamefont {Al{\`{u}}}},\ and\
  \bibinfo {author} {\bibfnamefont {E.}~\bibnamefont {Verhagen}},\ }\bibfield
  {title} {\bibinfo {title} {Nonreciprocity and magnetic-free isolation based
  on optomechanical interactions},\ }\bibfield  {journal} {\bibinfo  {journal}
  {Nature Communications}\ }\textbf {\bibinfo {volume} {7}},\ \href
  {https://doi.org/10.1038/ncomms13662} {10.1038/ncomms13662} (\bibinfo {year}
  {2016})\BibitemShut {NoStop}%
\bibitem [{\citenamefont {Li}\ \emph {et~al.}(2021)\citenamefont {Li},
  \citenamefont {Yin}, \citenamefont {Galiffi},\ and\ \citenamefont
  {Al\`u}}]{PhysRevLett.127.153903}%
  \BibitemOpen
  \bibfield  {author} {\bibinfo {author} {\bibfnamefont {H.}~\bibnamefont
  {Li}}, \bibinfo {author} {\bibfnamefont {S.}~\bibnamefont {Yin}}, \bibinfo
  {author} {\bibfnamefont {E.}~\bibnamefont {Galiffi}},\ and\ \bibinfo {author}
  {\bibfnamefont {A.}~\bibnamefont {Al\`u}},\ }\bibfield  {title} {\bibinfo
  {title} {Temporal parity-time symmetry for extreme energy transformations},\
  }\href {https://doi.org/10.1103/PhysRevLett.127.153903} {\bibfield  {journal}
  {\bibinfo  {journal} {Phys. Rev. Lett.}\ }\textbf {\bibinfo {volume} {127}},\
  \bibinfo {pages} {153903} (\bibinfo {year} {2021})}\BibitemShut {NoStop}%
\bibitem [{\citenamefont {Lustig}\ \emph {et~al.}(2018)\citenamefont {Lustig},
  \citenamefont {Sharabi},\ and\ \citenamefont {Segev}}]{Lustig18}%
  \BibitemOpen
  \bibfield  {author} {\bibinfo {author} {\bibfnamefont {E.}~\bibnamefont
  {Lustig}}, \bibinfo {author} {\bibfnamefont {Y.}~\bibnamefont {Sharabi}},\
  and\ \bibinfo {author} {\bibfnamefont {M.}~\bibnamefont {Segev}},\ }\bibfield
   {title} {\bibinfo {title} {Topological aspects of photonic time crystals},\
  }\href {https://doi.org/10.1364/OPTICA.5.001390} {\bibfield  {journal}
  {\bibinfo  {journal} {Optica}\ }\textbf {\bibinfo {volume} {5}},\ \bibinfo
  {pages} {1390} (\bibinfo {year} {2018})}\BibitemShut {NoStop}%
\bibitem [{\citenamefont {Wang}\ \emph {et~al.}(2021)\citenamefont {Wang},
  \citenamefont {Dutt}, \citenamefont {Wojcik},\ and\ \citenamefont
  {Fan}}]{Wang2021}%
  \BibitemOpen
  \bibfield  {author} {\bibinfo {author} {\bibfnamefont {K.}~\bibnamefont
  {Wang}}, \bibinfo {author} {\bibfnamefont {A.}~\bibnamefont {Dutt}}, \bibinfo
  {author} {\bibfnamefont {C.~C.}\ \bibnamefont {Wojcik}},\ and\ \bibinfo
  {author} {\bibfnamefont {S.}~\bibnamefont {Fan}},\ }\bibfield  {title}
  {\bibinfo {title} {Topological complex-energy braiding of non-hermitian
  bands},\ }\href {https://doi.org/10.1038/s41586-021-03848-x} {\bibfield
  {journal} {\bibinfo  {journal} {Nature}\ }\textbf {\bibinfo {volume} {598}},\
  \bibinfo {pages} {59} (\bibinfo {year} {2021})}\BibitemShut {NoStop}%
\bibitem [{\citenamefont {Joannopoulos}\ \emph {et~al.}(2008)\citenamefont
  {Joannopoulos}, \citenamefont {Johnson}, \citenamefont {Winn},\ and\
  \citenamefont {Meade}}]{Joannopoulos2008-mv}%
  \BibitemOpen
  \bibfield  {author} {\bibinfo {author} {\bibfnamefont {J.~D.}\ \bibnamefont
  {Joannopoulos}}, \bibinfo {author} {\bibfnamefont {S.~G.}\ \bibnamefont
  {Johnson}}, \bibinfo {author} {\bibfnamefont {J.~N.}\ \bibnamefont {Winn}},\
  and\ \bibinfo {author} {\bibfnamefont {R.~D.}\ \bibnamefont {Meade}},\
  }\href@noop {} {\emph {\bibinfo {title} {Photonic crystals}}},\ \bibinfo
  {edition} {2nd}\ ed.\ (\bibinfo  {publisher} {Princeton University Press},\
  \bibinfo {address} {Princeton, NJ},\ \bibinfo {year} {2008})\BibitemShut
  {NoStop}%
\bibitem [{\citenamefont {Sievenpiper}\ \emph {et~al.}(1999)\citenamefont
  {Sievenpiper}, \citenamefont {Zhang}, \citenamefont {Broas}, \citenamefont
  {g.~Alexopolous},\ and\ \citenamefont
  {Yablonovitch}}]{Sievenpiper1999HighimpedanceES}%
  \BibitemOpen
  \bibfield  {author} {\bibinfo {author} {\bibfnamefont {D.~F.}\ \bibnamefont
  {Sievenpiper}}, \bibinfo {author} {\bibfnamefont {L.}~\bibnamefont {Zhang}},
  \bibinfo {author} {\bibfnamefont {R.~J.}\ \bibnamefont {Broas}}, \bibinfo
  {author} {\bibfnamefont {N.}~\bibnamefont {g.~Alexopolous}},\ and\ \bibinfo
  {author} {\bibfnamefont {E.}~\bibnamefont {Yablonovitch}},\ }\bibfield
  {title} {\bibinfo {title} {High-impedance electromagnetic surfaces with a
  forbidden frequency band},\ }\href@noop {} {\bibfield  {journal} {\bibinfo
  {journal} {IEEE Transactions on Microwave Theory and Techniques}\ }\textbf
  {\bibinfo {volume} {47}},\ \bibinfo {pages} {2059} (\bibinfo {year}
  {1999})}\BibitemShut {NoStop}%
\bibitem [{\citenamefont {Galiffi}\ \emph {et~al.}(2022)\citenamefont
  {Galiffi}, \citenamefont {Tirole}, \citenamefont {Yin}, \citenamefont {Li},
  \citenamefont {Vezzoli}, \citenamefont {Huidobro}, \citenamefont
  {Silveirinha}, \citenamefont {Sapienza}, \citenamefont {Al{\`u}},\ and\
  \citenamefont {Pendry}}]{10.1117/1.AP.4.1.014002}%
  \BibitemOpen
  \bibfield  {author} {\bibinfo {author} {\bibfnamefont {E.}~\bibnamefont
  {Galiffi}}, \bibinfo {author} {\bibfnamefont {R.}~\bibnamefont {Tirole}},
  \bibinfo {author} {\bibfnamefont {S.}~\bibnamefont {Yin}}, \bibinfo {author}
  {\bibfnamefont {H.}~\bibnamefont {Li}}, \bibinfo {author} {\bibfnamefont
  {S.}~\bibnamefont {Vezzoli}}, \bibinfo {author} {\bibfnamefont {P.~A.}\
  \bibnamefont {Huidobro}}, \bibinfo {author} {\bibfnamefont {M.~G.}\
  \bibnamefont {Silveirinha}}, \bibinfo {author} {\bibfnamefont
  {R.}~\bibnamefont {Sapienza}}, \bibinfo {author} {\bibfnamefont
  {A.}~\bibnamefont {Al{\`u}}},\ and\ \bibinfo {author} {\bibfnamefont {J.~B.}\
  \bibnamefont {Pendry}},\ }\bibfield  {title} {\bibinfo {title} {{Photonics of
  time-varying media}},\ }\href {https://doi.org/10.1117/1.AP.4.1.014002}
  {\bibfield  {journal} {\bibinfo  {journal} {Advanced Photonics}\ }\textbf
  {\bibinfo {volume} {4}},\ \bibinfo {pages} {014002} (\bibinfo {year}
  {2022})}\BibitemShut {NoStop}%
\bibitem [{\citenamefont {Galiffi}\ \emph {et~al.}(2023)\citenamefont
  {Galiffi}, \citenamefont {Xu}, \citenamefont {Yin}, \citenamefont {Moussa},
  \citenamefont {Ra'di},\ and\ \citenamefont {Al{\`{u}}}}]{Galiffi2023}%
  \BibitemOpen
  \bibfield  {author} {\bibinfo {author} {\bibfnamefont {E.}~\bibnamefont
  {Galiffi}}, \bibinfo {author} {\bibfnamefont {G.}~\bibnamefont {Xu}},
  \bibinfo {author} {\bibfnamefont {S.}~\bibnamefont {Yin}}, \bibinfo {author}
  {\bibfnamefont {H.}~\bibnamefont {Moussa}}, \bibinfo {author} {\bibfnamefont
  {Y.}~\bibnamefont {Ra'di}},\ and\ \bibinfo {author} {\bibfnamefont
  {A.}~\bibnamefont {Al{\`{u}}}},\ }\bibfield  {title} {\bibinfo {title}
  {Broadband coherent wave control through photonic collisions at time
  interfaces},\ }\bibfield  {journal} {\bibinfo  {journal} {Nature Physics}\
  }\href {https://doi.org/10.1038/s41567-023-02165-6}
  {10.1038/s41567-023-02165-6} (\bibinfo {year} {2023})\BibitemShut {NoStop}%
\bibitem [{\citenamefont {Wang}\ \emph {et~al.}(2023)\citenamefont {Wang},
  \citenamefont {Mirmoosa}, \citenamefont {Asadchy}, \citenamefont {Rockstuhl},
  \citenamefont {Fan},\ and\ \citenamefont
  {Tretyakov}}]{doi:10.1126/sciadv.adg7541}%
  \BibitemOpen
  \bibfield  {author} {\bibinfo {author} {\bibfnamefont {X.}~\bibnamefont
  {Wang}}, \bibinfo {author} {\bibfnamefont {M.~S.}\ \bibnamefont {Mirmoosa}},
  \bibinfo {author} {\bibfnamefont {V.~S.}\ \bibnamefont {Asadchy}}, \bibinfo
  {author} {\bibfnamefont {C.}~\bibnamefont {Rockstuhl}}, \bibinfo {author}
  {\bibfnamefont {S.}~\bibnamefont {Fan}},\ and\ \bibinfo {author}
  {\bibfnamefont {S.~A.}\ \bibnamefont {Tretyakov}},\ }\bibfield  {title}
  {\bibinfo {title} {Metasurface-based realization of photonic time crystals},\
  }\href {https://doi.org/10.1126/sciadv.adg7541} {\bibfield  {journal}
  {\bibinfo  {journal} {Science Advances}\ }\textbf {\bibinfo {volume} {9}},\
  \bibinfo {pages} {eadg7541} (\bibinfo {year} {2023})}\BibitemShut {NoStop}%
\bibitem [{\citenamefont {Sounas}\ and\ \citenamefont
  {Al{\`{u}}}(2017)}]{Sounas2017}%
  \BibitemOpen
  \bibfield  {author} {\bibinfo {author} {\bibfnamefont {D.~L.}\ \bibnamefont
  {Sounas}}\ and\ \bibinfo {author} {\bibfnamefont {A.}~\bibnamefont
  {Al{\`{u}}}},\ }\bibfield  {title} {\bibinfo {title} {Non-reciprocal
  photonics based on time modulation},\ }\href
  {https://doi.org/10.1038/s41566-017-0051-x} {\bibfield  {journal} {\bibinfo
  {journal} {Nature Photonics}\ }\textbf {\bibinfo {volume} {11}},\ \bibinfo
  {pages} {774} (\bibinfo {year} {2017})}\BibitemShut {NoStop}%
\bibitem [{\citenamefont {Bhandare}\ \emph {et~al.}(2005)\citenamefont
  {Bhandare}, \citenamefont {Ibrahim}, \citenamefont {Sandel}, \citenamefont
  {Zhang}, \citenamefont {Wust},\ and\ \citenamefont {Noe}}]{1425478}%
  \BibitemOpen
  \bibfield  {author} {\bibinfo {author} {\bibfnamefont {S.}~\bibnamefont
  {Bhandare}}, \bibinfo {author} {\bibfnamefont {S.}~\bibnamefont {Ibrahim}},
  \bibinfo {author} {\bibfnamefont {D.}~\bibnamefont {Sandel}}, \bibinfo
  {author} {\bibfnamefont {H.}~\bibnamefont {Zhang}}, \bibinfo {author}
  {\bibfnamefont {F.}~\bibnamefont {Wust}},\ and\ \bibinfo {author}
  {\bibfnamefont {R.}~\bibnamefont {Noe}},\ }\bibfield  {title} {\bibinfo
  {title} {Novel nonmagnetic 30-db traveling-wave single-sideband optical
  isolator integrated in iii/v material},\ }\href
  {https://doi.org/10.1109/JSTQE.2005.845620} {\bibfield  {journal} {\bibinfo
  {journal} {IEEE Journal of Selected Topics in Quantum Electronics}\ }\textbf
  {\bibinfo {volume} {11}},\ \bibinfo {pages} {417} (\bibinfo {year}
  {2005})}\BibitemShut {NoStop}%
\bibitem [{\citenamefont {Hadad}\ \emph {et~al.}(2015)\citenamefont {Hadad},
  \citenamefont {Sounas},\ and\ \citenamefont {Alu}}]{Hadad2015}%
  \BibitemOpen
  \bibfield  {author} {\bibinfo {author} {\bibfnamefont {Y.}~\bibnamefont
  {Hadad}}, \bibinfo {author} {\bibfnamefont {D.~L.}\ \bibnamefont {Sounas}},\
  and\ \bibinfo {author} {\bibfnamefont {A.}~\bibnamefont {Alu}},\ }\bibfield
  {title} {\bibinfo {title} {Space-time gradient metasurfaces},\ }\bibfield
  {journal} {\bibinfo  {journal} {Physical Review B}\ }\textbf {\bibinfo
  {volume} {92}},\ \href {https://doi.org/10.1103/physrevb.92.100304}
  {10.1103/physrevb.92.100304} (\bibinfo {year} {2015})\BibitemShut {NoStop}%
\bibitem [{\citenamefont {Qin}\ \emph {et~al.}(2014)\citenamefont {Qin},
  \citenamefont {Xu},\ and\ \citenamefont {Wang}}]{6887369}%
  \BibitemOpen
  \bibfield  {author} {\bibinfo {author} {\bibfnamefont {S.}~\bibnamefont
  {Qin}}, \bibinfo {author} {\bibfnamefont {Q.}~\bibnamefont {Xu}},\ and\
  \bibinfo {author} {\bibfnamefont {Y.~E.}\ \bibnamefont {Wang}},\ }\bibfield
  {title} {\bibinfo {title} {Nonreciprocal components with distributedly
  modulated capacitors},\ }\href {https://doi.org/10.1109/TMTT.2014.2347935}
  {\bibfield  {journal} {\bibinfo  {journal} {IEEE Transactions on Microwave
  Theory and Techniques}\ }\textbf {\bibinfo {volume} {62}},\ \bibinfo {pages}
  {2260} (\bibinfo {year} {2014})}\BibitemShut {NoStop}%
\bibitem [{\citenamefont {He}\ \emph {et~al.}(2023)\citenamefont {He},
  \citenamefont {Zhang}, \citenamefont {Qi}, \citenamefont {Bo},\ and\
  \citenamefont {Li}}]{He2023}%
  \BibitemOpen
  \bibfield  {author} {\bibinfo {author} {\bibfnamefont {H.}~\bibnamefont
  {He}}, \bibinfo {author} {\bibfnamefont {S.}~\bibnamefont {Zhang}}, \bibinfo
  {author} {\bibfnamefont {J.}~\bibnamefont {Qi}}, \bibinfo {author}
  {\bibfnamefont {F.}~\bibnamefont {Bo}},\ and\ \bibinfo {author}
  {\bibfnamefont {H.}~\bibnamefont {Li}},\ }\bibfield  {title} {\bibinfo
  {title} {Faraday rotation in nonreciprocal photonic time-crystals},\
  }\bibfield  {journal} {\bibinfo  {journal} {Applied Physics Letters}\
  }\textbf {\bibinfo {volume} {122}},\ \href
  {https://doi.org/10.1063/5.0131818} {10.1063/5.0131818} (\bibinfo {year}
  {2023})\BibitemShut {NoStop}%
\bibitem [{\citenamefont {Elnaggar}\ and\ \citenamefont
  {Milford}(2020)}]{9063633}%
  \BibitemOpen
  \bibfield  {author} {\bibinfo {author} {\bibfnamefont {S.~Y.}\ \bibnamefont
  {Elnaggar}}\ and\ \bibinfo {author} {\bibfnamefont {G.~N.}\ \bibnamefont
  {Milford}},\ }\bibfield  {title} {\bibinfo {title} {Modeling space–time
  periodic structures with arbitrary unit cells using time periodic circuit
  theory},\ }\href {https://doi.org/10.1109/TAP.2020.2985712} {\bibfield
  {journal} {\bibinfo  {journal} {IEEE Transactions on Antennas and
  Propagation}\ }\textbf {\bibinfo {volume} {68}},\ \bibinfo {pages} {6636}
  (\bibinfo {year} {2020})}\BibitemShut {NoStop}%
\bibitem [{\citenamefont {Jayathurathnage}\ \emph {et~al.}(2021)\citenamefont
  {Jayathurathnage}, \citenamefont {Liu}, \citenamefont {Mirmoosa},
  \citenamefont {Wang}, \citenamefont {Fleury},\ and\ \citenamefont
  {Tretyakov}}]{Jayathurathnage2021}%
  \BibitemOpen
  \bibfield  {author} {\bibinfo {author} {\bibfnamefont {P.}~\bibnamefont
  {Jayathurathnage}}, \bibinfo {author} {\bibfnamefont {F.}~\bibnamefont
  {Liu}}, \bibinfo {author} {\bibfnamefont {M.~S.}\ \bibnamefont {Mirmoosa}},
  \bibinfo {author} {\bibfnamefont {X.}~\bibnamefont {Wang}}, \bibinfo {author}
  {\bibfnamefont {R.}~\bibnamefont {Fleury}},\ and\ \bibinfo {author}
  {\bibfnamefont {S.~A.}\ \bibnamefont {Tretyakov}},\ }\bibfield  {title}
  {\bibinfo {title} {Time-varying components for enhancing wireless transfer of
  power and information},\ }\bibfield  {journal} {\bibinfo  {journal} {Physical
  Review Applied}\ }\textbf {\bibinfo {volume} {16}},\ \href
  {https://doi.org/10.1103/physrevapplied.16.014017}
  {10.1103/physrevapplied.16.014017} (\bibinfo {year} {2021})\BibitemShut
  {NoStop}%
\bibitem [{\citenamefont {Elnaggar}\ and\ \citenamefont
  {Milford}(2021)}]{Elnaggar2021}%
  \BibitemOpen
  \bibfield  {author} {\bibinfo {author} {\bibfnamefont {S.~Y.}\ \bibnamefont
  {Elnaggar}}\ and\ \bibinfo {author} {\bibfnamefont {G.~N.}\ \bibnamefont
  {Milford}},\ }\bibfield  {title} {\bibinfo {title} {Properties of translation
  operator and the solution of the eigenvalue and boundary value problems of
  arbitrary space{\textendash}time periodic metamaterials},\ }\href
  {https://doi.org/10.1098/rsos.210367} {\bibfield  {journal} {\bibinfo
  {journal} {Royal Society Open Science}\ }\textbf {\bibinfo {volume} {8}},\
  \bibinfo {pages} {210367} (\bibinfo {year} {2021})}\BibitemShut {NoStop}%
\bibitem [{\citenamefont {Bergholtz}\ \emph {et~al.}(2021)\citenamefont
  {Bergholtz}, \citenamefont {Budich},\ and\ \citenamefont
  {Kunst}}]{RevModPhys.93.015005}%
  \BibitemOpen
  \bibfield  {author} {\bibinfo {author} {\bibfnamefont {E.~J.}\ \bibnamefont
  {Bergholtz}}, \bibinfo {author} {\bibfnamefont {J.~C.}\ \bibnamefont
  {Budich}},\ and\ \bibinfo {author} {\bibfnamefont {F.~K.}\ \bibnamefont
  {Kunst}},\ }\bibfield  {title} {\bibinfo {title} {Exceptional topology of
  non-hermitian systems},\ }\href
  {https://doi.org/10.1103/RevModPhys.93.015005} {\bibfield  {journal}
  {\bibinfo  {journal} {Rev. Mod. Phys.}\ }\textbf {\bibinfo {volume} {93}},\
  \bibinfo {pages} {015005} (\bibinfo {year} {2021})}\BibitemShut {NoStop}%
\bibitem [{\citenamefont {Miri}\ and\ \citenamefont
  {Al{\`{u}}}(2019)}]{Miri2019}%
  \BibitemOpen
  \bibfield  {author} {\bibinfo {author} {\bibfnamefont {M.-A.}\ \bibnamefont
  {Miri}}\ and\ \bibinfo {author} {\bibfnamefont {A.}~\bibnamefont
  {Al{\`{u}}}},\ }\bibfield  {title} {\bibinfo {title} {Exceptional points in
  optics and photonics},\ }\bibfield  {journal} {\bibinfo  {journal} {Science}\
  }\textbf {\bibinfo {volume} {363}},\ \href
  {https://doi.org/10.1126/science.aar7709} {10.1126/science.aar7709} (\bibinfo
  {year} {2019})\BibitemShut {NoStop}%
\end{thebibliography}%
\end{document}